\documentclass[12pt]{iopart_mod}

\usepackage{amsfonts,amssymb,amsmath,amsthm,epsfig}
\usepackage{cite}

\newcommand{\eps}{\varepsilon}

\newcommand{\beqs}{\begin{equation*}}
\newcommand{\beq}{\begin{equation}}

\newcommand{\eeqs}{\end{equation*}}
\newcommand{\eeq}{\end{equation}}

\newcommand{\beqas}{\begin{eqnarray*}}
\newcommand{\beqa}{\begin{eqnarray}}

\newcommand{\eeqas}{\end{eqnarray*}}
\newcommand{\eeqa}{\end{eqnarray}}

\newcommand{\eq}[2]{\begin{equation} #1 \label{#2} \end{equation}}

\newcommand{\blist}{\begin{itemize}}

\newcommand{\elist}{\end{itemize}}

\providecommand{\href}[2]{#2}

\DeclareMathOperator{\extdm}{d}
\newcommand{\extd}{\extdm \!}

\newcommand{\newpot}{W}
\newcommand{\Lambert}{W_0}

\begin{document}


\hfill {\footnotesize MIT-CTP~3842}

{\hfill {\tt \footnotesize arXiv:0706.4070}}

\title{Poisson-sigma model for 2D gravity with non-metricity}

\author{M.~Adak}

\address{Department of Physics, Faculty of Arts and Sciences, 
 Pamukkale University, \\ 20017 Denizli, Turkey}

\author{D.~Grumiller}

\address{Center for Theoretical Physics,
Massachusetts Institute of Technology,\\
77 Massachusetts Ave.,
Cambridge, MA  02139}

\eads{\mailto{madak@pau.edu.tr}, \mailto{grumil@lns.mit.edu}}

\begin{abstract}
We present a Poisson-sigma model describing general 2D dilaton gravity with non-metricity, torsion and curvature. 
It involves three arbitrary functions of the dilaton field, two of which are well-known from metric compatible theories, while the third one characterizes the local strength of non-metricity. 
As an example we show that $\alpha^\prime$ corrections in 2D string theory can generate (target space) non-metricity.
\end{abstract}



\setcounter{footnote}{0}

\section{Introduction}\label{sec:1}
    
Dilaton gravity in two dimensions has found many classical, semi-classical and quantum applications \cite{Grumiller:2002nm}. A very efficient way to describe it and to exhibit integrability is by means of a first order formulation \cite{Isler:1989hq} 
in terms of Cartan variables, dyad $e^a_\mu$ and connection $\omega^a{}_{b\,\mu}$. It is closely related to a specific type of non-linear gauge theory \cite{Ikeda:1994fh}, a Poisson-sigma model (PSM) \cite{Schaller:1994es}. However, it needs some extra structure beyond the one provided by the PSM, namely the specification of a tangent space metric $\eta_{ab}$. The metric then follows uniquely from this specification, $g_{\mu\nu}=e^a_\mu e^b_\nu \eta_{ab}$, where the fields $e^a_\mu$ are determined from the dynamics of the PSM. 
In \cite{Strobl:2003kb} it was suggested to introduce the tangent space metric as an external tensor, whose compatibility with the PSM structure was analyzed carefully. All relevant examples presented there introduce $\eta_{ab}$ as a fixed metric depending on certain 0-form fields. Thus $\eta_{ab}$ is not considered as an independent field. It would be nice if there was a pure PSM formulation that allowed to introduce the tangent space metric as an independent field. This is the first problem we would like to address.
A related drawback of the standard PSM formulation is the difficulty to incorporate non-metricity. So far only isolated second order models are known (or equivalent first order formulations that are not manifestly a PSM) \cite{Dereli:1994vx,Obukhov:2003tz,Adak:2005ns}, which considerably complicates their classical (and quantum) analysis. 
This is the second problem we would like to address. In fact, we shall demonstrate in this Letter that both problems can be solved in a simple way. 
Namely, we propose to employ the first order action ($U$, $V$ and $W$ are some functions defining the model)
\begin{multline}
\label{eq:action}
I=k\int_{\mathcal{M}_2} \Big[X\extd\omega + X^a \left(\delta_a^b \extd + \epsilon_a{}^b\omega\right)\wedge e_b + 
\rho_{ab} \wedge  \big(\extd\eta^{ab} - W(X)\eta^{ab} X^c\epsilon_c{}^de_d\big)  \\
+ \frac12 \left(V(X) \eta^{ac} + U(X) X^aX^c\right)\epsilon_c{}^b\,e_b\wedge e_a\Big]\,.
\end{multline}
We shall prove in Section \ref{sec:2} that \eqref{eq:action} is a specific PSM.  Our notation is explained there as well. Section \ref{sec:3} provides a derivation of all classical solutions descending from \eqref{eq:action}. The final Section \ref{sec:4} contains a discussion, an application and a comparison with existing special cases of two-dimensional dilaton gravity models with (or without) non-metricity. 

\section{Action}\label{sec:2}

Let us consider a PSM \cite{Schaller:1994es} (we drop all boundary terms in this work)
\eq{
I=k\int_{\mathcal{M}_2} \Big[ A_I \extd X^I - \frac12 P^{IJ} A_J\wedge A_I\Big]=k\int_{\mathcal{M}_2} \Big[X^I\extd A_I - \frac12 P^{IJ} A_J\wedge A_I\Big]\,,
}{eq:nm1}
with the field content $X^I=(X,X^a,\eta^{ab})$ and $A_I=(\omega,e_a,\rho_{ab})$. The former are the target space coordinates of a Poisson manifold with the Poisson tensor $P^{IJ}$ depending on them, while the latter are gauge field 1-forms, e.g.~$e_a=e_{a\,\mu}\extd x^\mu$. Here the index $a$ runs over two values, which we denote by $+,-$, while the index $ab$ runs over three values, which we denote by $++,+-,--$. For convenience we also introduce the value $-+$ for $ab$ and assume $\eta^{ab}=\eta^{ba}$ and $\rho_{ab}=\rho_{ba}$. This allows us to consider the index $ab$ as an index pair. The indices $I,J,\dots$ always run over the full set $X,+,-,++,+-,--$. The index position was chosen for sake of similarity with the PSM notation, but one could equally define target space coordinates of the form $X_a$, $\eta_{ab}$ and gauge fields $e^a$, $\rho^{ab}$. The construction of a metric $g=e^ae^b\eta_{ab}$ or $g=e_ae_b\eta^{ab}$ requires that $X^a$ and $\eta^{ab}$ have the {\em same} index positions. A priori there is no device to manipulate indices, besides the Levi-Civita and Kronecker symbols. The real coupling constant $k$ is irrelevant for our discussion.
The suggestive notation is chosen for sake of clarity, but we emphasize that at this point the fields do not have some geometric interpretation in terms of ``dilaton field'', ``Zweibeine'' or ``connection''. Neither do the indices $\pm$ necessarily imply light cone variables, the use of which is very convenient in two-dimensional gravity \cite{Polyakov:1987zb,Kummer:1992bg}. We employ the Einstein summation convention and introduce the abbreviation
\eq{
\epsilon_a{}^b := \tilde{\epsilon}_{ac} \eta^{cb}
}{eq:nm2}
with the Levi-Civita symbol $\tilde{\epsilon}_{ab}=-\tilde{\epsilon}_{ba}$, where  $\tilde{\epsilon}_{-+} = 1$. From these definitions we derive the useful relations $\epsilon_a{}^a=0$ and $\epsilon_a{}^c\epsilon_c{}^b=(-\det{\eta})\delta_a^b$,
where $\det{\eta}:=\eta^{++}\eta^{--}-\eta^{+-}\eta^{+-}$ and $\delta_a^b$ is the Kronecker symbol. For the Poisson tensor $P^{IJ}=-P^{JI}$ we choose\footnote{For transparency we introduce a comma between first and second index.}
\begin{align}
P^{X,a} &= X^b\epsilon_b{}^a \label{eq:P1} \\
P^{a,b} &= -\frac{1}{2} V(X) \left(\eta^{ac}\epsilon_c{}^b-\eta^{bc}\epsilon_c{}^a\right) - \frac{1}{2} U(X)\left(X^aX^c\epsilon_c{}^b-X^bX^c\epsilon_c{}^a\right) \label{eq:P2} \\
P^{X,ab} &= 0 \label{eq:P3} \\
P^{ab,cd} &= 0 \label{eq:P4} \\
P^{a,bc} &= \newpot(X) X^d \epsilon_d{}^a\eta^{bc} 
\label{eq:P5}
\end{align}
We shall comment on the specific form of $P^{IJ}$ in Section \ref{sec:4} and confine ourselves to a couple of immediate remarks. The choices \eqref{eq:P1}-\eqref{eq:P2} are standard \cite{Schaller:1994es,Grumiller:2002nm}, while \eqref{eq:P3}-\eqref{eq:P4} were chosen for simplicity. With \eqref{eq:P1}-\eqref{eq:P4} as Ansatz the non-trivial entry \eqref{eq:P5} 
emerges as solution of the non-linear Jacobi identities
\eq{
J^{IJK}:= P^{IL} \partial_L P^{JK} + P^{JL} \partial_L P^{KI} +  P^{KL} \partial_L P^{IJ} = 0\,.
}{eq:nm4}
Only if they are fulfilled then $P^{IJ}$ as defined in \eqref{eq:P1}-\eqref{eq:P5} really is a Poisson tensor and the action \eqref{eq:nm1} is a PSM [or, equivalently up to a surface term, the action \eqref{eq:action}].

We check now the validity of \eqref{eq:nm4}. To reduce clutter we shall immediately drop all terms containing $P^{X,ab}$ or $P^{ab,cd}$ or derivatives thereof. Decomposing the generic indices $I,J,\dots$ into $X,a,ab$ the identities we have to check split into $J^{X,a,b}=J^{X,a,bc}=J^{X,ab,cd}=J^{a,b,cd}=J^{a,bc,de}=J^{ab,cd,ef}=0$. Let us start with the first one: Without loss of generality we can simplify $J^{X,a,b}=0$ to $J^{X,+,-}=0$ and get
\begin{equation}
P^{X,c}\partial_cP^{+,-}+P^{+,-}\partial_aP^{a,X}+P^{+,cd}\partial_{cd} P^{-,X}+P^{-,cd}\partial_{cd} P^{X,+}=0\,.
\label{eq:nm6}
\end{equation}
The first term $P^{X,c}\partial_c P^{+,-}\propto X^b\epsilon_b{}^c\partial_c (X^-X^d\epsilon_d{}^+-X^+X^d\epsilon_d{}^-) = 0$ vanishes by itself. The second term vanishes because $\partial_aP^{a,X}=0$. Therefore we obtain the condition
\eq{
X^-(P^{+,+-}-P^{-,++})=X^+(P^{+,--}-P^{-,+-})\,,
}{eq:nm12}
which is fulfilled for \eqref{eq:P5}. The second identity $J^{X,a,bc}=0$,
\begin{equation}
P^{X,d}\partial_d P^{a,bc} = P^{d,bc}\partial_d P^{X,a} \,,
\label{eq:nm7}
\end{equation}
holds because \eqref{eq:P5} has a structure very similar to \eqref{eq:P1}. The third identity $J^{X,ab,cd}=0$ holds identically because $P^{X,ab}$ and $P^{ab,cd}$ vanish. The fourth identity $J^{a,b,cd}=0$ splits into three parts,
\eq{
P^{a,I}\partial_I P^{b,cd} - P^{b,I}\partial_I P^{a,cd} = P^{e,cd}\partial_e P^{a,b} \,.
}{eq:nm13}
The terms on the left hand side actually cancel each other, so the right hand side must vanish by itself, which indeed is the case: $P^{e,cd}\partial_e P^{a,b}=-1/2\,U\newpot \eta^{cd} X^f\epsilon_f{}^e\partial_e(X^aX^g\epsilon_g{}^b-X^bX^g\epsilon_g{}^a)=0$. The fifth identity $J^{a,bc,de}=0$ simplifies to
\eq{
P^{f,bc}\partial_f P^{a,de} = P^{f,de}\partial_f P^{a,bc}\,.
}{eq:nm14}
Evidently this relation holds for \eqref{eq:P5}. The sixth and final identity, $J^{ab,cd,ef}=0$, is trivially fulfilled because of the choice \eqref{eq:P4}. Thus we conclude that \eqref{eq:P1}-\eqref{eq:P5} is a valid Poisson tensor because all Jacobi identities \eqref{eq:nm4} hold.

\section{Classical solutions}\label{sec:3}

The gauge symmetries \cite{Ikeda:1994fh} of the PSM action \eqref{eq:nm1},
\eq{
\delta_\eps X^I=P^{IJ}\eps_J\,,\qquad \delta_\eps A_I=-\extd\eps_I-\big(\partial_IP^{JK}\big)\eps_K A_J\,,
}{eq:nm15}
render the model a topological one in the sense that there are no propagating physical degrees of freedom \cite{Schaller:1994es}. The integration of the equations of motion,
\eq{
\extd X^I+P^{IJ}A_J=0\,,\qquad\extd A_I - \frac12 \big(\partial_I P^{JK}\big) A_K\wedge A_J = 0\,,
}{eq:nm16}
leads to four conserved quantities in the present case. This can be deduced from the form of the Poisson-tensor \eqref{eq:P1}-\eqref{eq:P5}: its dimension is six, but its rank equals to two, so there is a four-dimensional kernel corresponding to the four conserved quantities, also known as Casimir functions. We demonstrate now explicitly how to obtain them using the equations of motion \eqref{eq:nm16}. We shall need only the first set of equations ($\hat{X}^a:=X^b \epsilon_b{}^a$),
\begin{align}
\extd X &= -\hat{X}^a e_a\,,\label{eq:eom1} \\
\extd X^a &= \hat{X}^a\omega - P^{a,b}e_b - \newpot(X)\hat{X}^a \eta^{bc}\rho_{bc} \,,\label{eq:eom2} \\
\extd \eta^{ab} &= \newpot(X)\hat{X}^ce_c \eta^{ab}\,.\label{eq:eom3}
\end{align}
The first and third equation lead to three conserved quantities $\eta^{ab}_{(0)}$,
\eq{
\extd\left(e^{P(X)}\eta^{ab}\right)=0\qquad \Rightarrow \qquad \eta^{ab}=\eta^{ab}_{(0)} \,e^{-P(X)}\,,
}{eq:casimirnew}
with
\eq{
P(X):=\int^X \!\!\extd y\,\newpot(y)\,.
}{eq:P}
It is now obvious that the function $\newpot$ is responsible for non-metricity because $\extd \eta^{ab}$ vanishes if $\newpot$ vanishes. 
Introducing
\eq{
Y:=X^a\hat{X}^b\tilde{\epsilon}_{ba} = X^aX^b\epsilon_b{}^c\tilde{\epsilon}_{ca}
}{eq:Y}
and manipulating \eqref{eq:eom2} [inserting repeatedly \eqref{eq:eom1} and exploiting \eqref{eq:eom3} to simplify $\extd\hat{X}^a=\epsilon_b{}^a\extd X^b+\newpot\hat{X}^a\hat{X}^be_b$] yields
\eq{
\extd Y = -2\hat{X}^a\tilde{\epsilon}_{ab}P^{b,c}e_c-\newpot Y\extd X=\big[2V\det{\eta} - (U+\newpot)Y\big]\extd X\,.
}{eq:nm17}
The last equation together with \eqref{eq:casimirnew} establishes the fourth conserved quantity $M$,
\eq{
\extd\left(e^{Q(X)+P(X)}\,Y+w(X)\right)=0\qquad \Rightarrow \qquad e^{Q(X)+P(X)}\,Y+w(X) = M\,.
}{eq:casimir}
Here we have defined
\eq{
Q(X):=\int^X\!\!\extd y \,U(y)\,,\qquad w(X):= -2 \det{\eta_{(0)}} \int^X\!\!\extd y \, e^{Q(y)-P(y)} V(y)\,.
}{eq:Qw}
We would like to comment on the integration constants inherent to the definitions \eqref{eq:P} and \eqref{eq:Qw}. Obviously, a constant shift in $P$ is of no relevance and can be absorbed into a redefinition of $\eta^{ab}_{(0)}$. A constant shift of $Q$ corresponds to a rescaling of the physical units for length and mass, a well-known feature of the metric-compatible case where the same issue arises \cite{Grumiller:2002nm,Grumiller:2006rc}. A constant shift of $w$ can be absorbed into a redefinition of $M$.

Much like in the metric-compatible case \cite{Grumiller:2002nm}, $M$ has an interpretation as ``mass''. We show this by constructing the metric
\eq{
g_{\mu\nu} = e_{a\,\mu} e_{b\,\nu} \eta^{ab}_{(0)}e^{-P(X)}\,.
}{eq:nm20}
To simplify the discussion we shall assume\footnote{Other choices for $\eta^{ab}_{(0)}$ would lead to a similar discussion provided $\det\eta_{(0)}\neq 0$.} $\eta^{\pm\mp}_{(0)}=1$ and $\eta^{\pm\pm}_{(0)}=0$, which establishes $\epsilon_+{}^+=-\epsilon_-{}^-=-e^{-P}$ and $Y=2e^{-P}X^+X^-$.  A simple class of solutions is obtained for $X^\pm=0$, which by virtue of \eqref{eq:eom1} leads to constant $X$. Equation \eqref{eq:eom2} then requires that $X$ be a solution of $V(X)=0$ and \eqref{eq:eom3} establishes metricity, $\extd\eta^{ab}=0$. These solutions, so-called ``constant dilaton vacua'', are therefore the same as in the metric compatible case and allow only for constant curvature solutions, cf.~e.g.~\cite{Grumiller:2006rc}. They are non-generic because $V(X)=0$ need not have any solution in the range of definition of $X$. We shall now discuss the generic class of solutions, which requires $X\neq \rm const.$ Following \cite{Klosch:1996fi,Grumiller:2002nm} we make the Ansatz $e_-=X^+ Z$, where $Z$ is a 1-form. This Ansatz is valid in a patch where $X^+\neq 0$. Then \eqref{eq:eom1} yields $e_+=e^P \extd X/X^++X^-Z$, while \eqref{eq:eom2} yields $\omega-\newpot\eta^{ab}\rho_{ab}=-e^P\extd X^+/X^+ - e^P P^{+,-}Z$. Because of redundancy only one of the right equations \eqref{eq:nm16}, 
\eq{
\extd e_- + e^{-P}(\omega-\newpot\eta^{ab}\rho_{ab})\wedge e_- = \partial_- P^{+,-}e_-\wedge e_+\,,
}{eq:nm21}
is needed. Defining $Z=e^Q\hat{Z}$ and using the previous equations simplifies \eqref{eq:nm21} to $\extd\hat{Z}=0$. Since $\hat{Z}$ is closed, locally it is exact, $\hat{Z}=\extd u$. This yields $e_-=X^+e^Q\extd u$ and $e_+=e^P \extd X/X^++X^-e^Q\extd u$. Using $X$ and $u$ as coordinates we obtain from \eqref{eq:casimir} and \eqref{eq:nm20} the line-element
\eq{
\extd s^2 = g_{\mu\nu}\extd x^\mu\extd x^\nu = e^{Q(X)}\left[2\extd X\extd u + e^{-P(X)}\left(M-w(X)\right)\extd u^2\right]
}{eq:nm22}
in Eddington-Finkelstein gauge. Besides the ``constant dilaton vacua'' above, this is the most general classical solution for the line-element in an Eddington-Finkelstein patch. It is parameterized by a single constant of motion $M$ and exhibits a Killing vector $\partial_u$. For $P=0$ equation \eqref{eq:nm22} agrees with results of the metric-compatible case \cite{Klosch:1996fi,Grumiller:2002nm}. 

On a sidenote we mention that it is possible to obtain the same classical solutions \eqref{eq:nm22} also from a Riemannian second order action, 
\begin{equation}
\label{eq:actionRiemann}
I_2\propto\int_{\mathcal{M}_2} \!\! \extd^2x\sqrt{-g}\Big[\tilde{X} R  + \tilde{U}(\tilde{X}) (\nabla\tilde{X})^2 - 2\tilde{V}(\tilde{X})\Big] \,, 
\end{equation}
where $\extd\tilde{X}=e^{P(X)}\extd X$ and
\eq{
\tilde{U}(\tilde{X}) = (U(X)-W(X))e^{-P(X)}\,,\quad \tilde{V}(\tilde{X})=V(X)e^{-P(X)}\,.
}{eq:angelinajolie}
The reformulation \eqref{eq:actionRiemann}, \eqref{eq:angelinajolie} is useful for deriving thermodynamical properties (after adding appropriate boundary terms) \cite{Grumiller:2007ju}. One remarkable consequence of non-metricity is that the black hole entropy no longer is proportional to the original dilaton $X$, but rather to $\tilde{X}$.

\section{Discussion}\label{sec:4}

We demonstrated that \eqref{eq:P1}-\eqref{eq:P5} is a valid Poisson tensor for a PSM \eqref{eq:nm1} which allows an interpretation as a first-order gravity system with non-metricity \eqref{eq:action}. We then constructed all classical solutions for the line-element in a basic Eddington-Finkelstein patch \eqref{eq:nm22}. We did not address global properties, but such a discussion can be performed in analogy\footnote{It would be interesting to see the effect of $W\neq 0$ on global properties. Singularities of the non-metricity potential $W$ typically do not change the number and types of Killing horizons, but they can be of relevance for the asymptotic structure of space-time or geodesic (in)completeness properties. Moreover, the fact that we doubled the number of target space coordinates as compared to previous PSM approaches may have an impact on global considerations.} to \cite{Klosch:1996fi,Grumiller:2002nm}. For vanishing non-metricity potential, $W=0$, we recovered well-known results. 

Actually the last point can be seen already at the level of the action. Consider \eqref{eq:action} with $\newpot=0$. Then the $\eta,\rho$ sector decouples and can be integrated out trivially, leading to three Casimirs $\eta^{ab}=\eta^{ab}_{(0)}$. Let us choose them as $\eta^{\pm\pm}=0$ and $\eta^{\pm\mp}=1$. Then \eqref{eq:nm2} simplifies to $\epsilon_\mp{}^\mp=\pm 1$ 
and the action \eqref{eq:action} simplifies to
\eq{
I=k\int_{\mathcal{M}_2}\!\! \Big[X\extd \omega + X^\pm \left(\extd \mp \omega \right) \wedge e_\pm  +  \left(V(X) + X^+X^-U(X)\right) e_-\wedge e_+\Big]\,.
}{eq:nm5}
This coincides with the PSM action for ordinary dilaton gravity, parameterized by the potentials $U$ and $V$.
For instance, with $k=-1$ and $e^\pm = e_\mp$ the result \eqref{eq:nm5} coincides with Eq.~(2.2)
in \cite{Grumiller:2006rc}. So the simple case $\newpot=0$ is well understood.
The conceptual difference to previous approaches is that the tangent space metric $\eta^{ab}$ here is not
an external input but rather emerges from the integration of \eqref{eq:eom3}.
By choosing different values for the constants of motion $\eta^{ab}_{(0)}$ we can obtain either signature 
of the tangent space metric. Thus, the choice of signature in our approach happens only at the level of the equations of motion and not at the level of the action. 

We discuss now in a bit more detail the specific form \eqref{eq:P1}-\eqref{eq:P5} of the Poisson tensor and the geometric interpretation of our fields. The entry \eqref{eq:P1} is basically fixed by requiring that the torsion 2-form\footnote{Because we demand compatibility with the PSM index structure quantities like torsion and curvature have non-standard index positions of the tangent space indices.} 
\eq{
T_a := \extd e_a + \omega_a{}^b \wedge e_b = \extd e_a + \epsilon_a{}^b \omega\wedge e_b
}{eq:torsion}
appears in the actions \eqref{eq:action} and \eqref{eq:nm1}.
This interpretation, however, requires
anti-symmetry of the connection $\omega_a{}^b$. We achieve this
via \eqref{eq:P3} which eliminates any coupling of the connection
different from \eqref{eq:torsion}. Since this does not seem to be
the usual way of introducing non-metricity \cite{Hehl:1995ue} we
feel obliged to explain this point. Traditionally non-metricity,
 \eq{
  Q^{ab}:= \extd \eta^{ab} + \Lambda^{ab} + \Lambda^{ba}
}{eq:nonmetricity}
 requires a symmetric contribution to the full
connection, $\Lambda^{ab}\neq-\Lambda^{ba}$, because normally the
term $\extd\eta^{ab}$ vanishes since the tangent space metric is
assumed to be constant. However, rather than shifting the
burden of non-metricity to the symmetric part of the full 
connection one can achieve
non-metricity also with an anti-symmetric connection
$\omega_a{}^b$ by choosing a tangent space metric which is not
constant. A reasonable non-trivial choice for $\eta^{ab}$ in
physical applications could be an $(A)dS$ metric, for instance.
This re-interpretation of non-metricity is not tied to our
two-dimensional discussion but generalizes to higher dimensions.
For sake of completeness we mention that the full curvature
\eq{
R_a{}^b=\extd\omega_a{}^b=\epsilon_a{}^b\extd\omega - \tilde{\epsilon}_{ac}\,\omega\wedge\extd\eta^{cb}
}{eq:curvature}
is in general inequivalent to the Riemannian curvature derived from our solution for the line-element \eqref{eq:nm22}. Solely for
$U=W=0$ both notions of curvature agree with each other, because
then torsion and non-metricity vanish. After this digression we
return to the discussion of the Poisson tensor. The choice
\eqref{eq:P2} appears to be the most general expression compatible
with the index structure, anti-symmetry and the Jacobi-identities.
A novel feature 
is that \eqref{eq:P2} contains terms at most quadratic in $X^a$, whereas
in the traditional approach arbitrary coupling to
$X^aX^b\eta_{ab}$ is allowed since $\eta_{ab}$ is introduced there
as an external structure. This is interesting by itself because
supergravity imposes a similar restriction to quadratic coupling
\cite{Bergamin:2002ju}, but we shall not pursue this issue any
further here. As mentioned before the choice \eqref{eq:P4} was
made for simplicity. However, the Jacobi-identities \eqref{eq:nm4}
are very restrictive concerning contributions to \eqref{eq:P4} and
it could well be they imply $P^{ab,cd}=0$. The main purpose of our
Poisson tensor is to produce non-metricity, \eq{ \extd\eta^{bc} =
e_a f^{abc}\,, }{eq:nm23} where $f^{abc}=f^{acb}$. This is the
generic form of non-metricity because we can decompose any 1-form
into basis 1-forms $e_a$. The equations of motion \eqref{eq:nm16}
imply $f^{abc}=P^{a,bc}$ if $P^{X,ab}=P^{ab,cd}=0$. Thus, it is
natural to introduce a non-vanishing $P^{a,bc}$ component, but no
$P^{ab,cd}$ component is required to generate non-metricity. Our
choice \eqref{eq:P5} contains one free function parameterizing the
strength of non-metricity and appears to be the generic solution
of the Jacobi-identities \eqref{eq:nm4} once the Ansatz
\eqref{eq:P1}-\eqref{eq:P4} is taken for granted. It is evident
from \eqref{eq:eom3} that we have only a trace part of
non-metricity. So either a PSM formulation does not allow for a tracefree contribution to non-metricity
or it requires a different choice of the Poisson tensor.
Therefore it would be interesting to check in what sense our
choice \eqref{eq:P1}-\eqref{eq:P5} is generic, possibly by
adapting the discussion in \cite{Strobl:2003kb}.

It is worthwhile mentioning that the symmetries \eqref{eq:nm15} are non-linear because the Poisson tensor \eqref{eq:P1}-\eqref{eq:P5} is at least quadratic in the target space coordinates. Hence the quantities $\partial_I P^{JK}$ become structure functions rather than structure constants. Even for the metric-compatible case $W=0$ this differs from the simpler situation in \eqref{eq:nm5}: there the Jackiw-Teitelboim model $V\propto X$, $U=0$ \cite{JT} simplifies to a $SO(1,2)$ gauge theory. The reason for this difference comes from our treatment of $\eta^{ab}$ as target space coordinates.

Our main result derived from the action is the line-element \eqref{eq:nm22}, which depends on one additional function $P(X)$ as compared to standard results. This additional freedom can be useful in ``reverse-engineering procedures'' where one attempts to construct an action for a given family of classical solutions for dilaton field and line-element. For instance, it provides a new possibility to evade the no-go argument of \cite{Grumiller:2002md} and to construct a PSM action for the exact string black hole \cite{Dijkgraaf:1992ba} which differs from the one constructed in \cite{Grumiller:2005sq}.
A possible choice for the exact string black hole potentials is
\eq{
U=-\frac{X}{X^2+1}\,,\quad V=-\lambda X\,\frac{1+\sqrt{X^2+1}}{\sqrt{X^2+1}}\,,\quad W=\frac{U}{1+\sqrt{X^2+1}}\,.
}{eq:ESBH}
In this way the $\alpha^\prime$ corrections contained in the exact string black hole are encoded in non-metricity. In the weak coupling limit ($X\to\infty$) non-metricity becomes irrelevant ($P\to 1/X$) and the model asymptotes to the Witten black hole \cite{Witten:1991yr}.

Finally, we would like to comment on the relation to previous approaches. 
The action \eqref{eq:action} contains as special cases all models with non-metricity constructed so far. In particular, the results of \cite{Adak:2005ns} are recovered for 
\begin{align}
\label{eq:potantials}  
U &= - e^{-1-\Lambert(z)} \left[\frac{A}{1+\Lambert(z)}+\frac{B}{(1+\Lambert(z))^2}\right]\,, & \quad V & = \frac{2\beta}{k^2} e^{-1-\Lambert(z)} (1+\Lambert(z))^2 \,, \nonumber \\
W &=-\frac{k^2}{8}\, \frac{e^{-1-\Lambert(z)}}{(1+\Lambert(z))}\,, & \quad z &:=\frac{k^2 X}{8e} - \frac 1e \,.
\end{align}
The real parameters $A=(k+p)k/2 + k^2/8$ and $B=k^2(\alpha+\mu k^2+\nu(p^2-q^2))/16$ are related to parameters defined in that work ($k,
p,q,\alpha,\beta,\mu,\nu$). For brevity we have set an additional parameter to zero, $l=0$, but also the case $l\neq 0$ allows a comparison and leads to somewhat lengthy expressions for $U,V$ and $W$.
 The function $\Lambert$ denotes the principal branch of the Lambert-W function \cite{Corless:1996}. 
It is convenient to choose the free integration constant in $P$ such that $P=-1-\Lambert(z)$ (since $X$ typically is non-negative $z\geq -1/e$ and $\Lambert(z)\geq -1$, so that $P\leq 0$). 
The results of \cite{Dereli:1994vx} are a special case of \cite{Adak:2005ns} for $p=q=\nu=0$. The results of \cite{Obukhov:2003tz} are recovered for 
\begin{equation}
\label{eq:potantials1}  U(X) = B\tilde{X}\,,\qquad V(X) = C\tilde{X}+D\tilde{X}^2+E\tilde{X}^3 \,,\qquad W(X)=A\,. 
\end{equation}
The real parameters $A=a_1/a$, $B=a_1^2/a^2$, $C=\lambda/a+D/a_1$, $D=a_1(4a_2-1)/(2ab)$ and $E=-a_1^2/(4ab)$ are related to parameters defined in that work ($a,a_1,a_2,b,\lambda$). The dilaton $X$ is determined from $\tilde{X}$ by the relation between the equations \eqref{eq:actionRiemann} and \eqref{eq:angelinajolie}, which integrates to $A X=\ln{(A\tilde{X})}$.
Our action \eqref{eq:action} not only encompasses all these special cases, but generalizes them while maintaining integrability. 
It could be interesting to couple matter to the system, to supersymmetrize it and/or to quantize it. This should be possible by analogy to the metric compatible case \cite{Grumiller:2002nm}, even though there will be technical and conceptual differences because the tangent-space metric $\eta^{ab}$ now is dynamical.

\section*{Acknowledgments}

We thank Tekin Dereli, Roman Jackiw, Yuri Obukhov and especially Thomas Strobl for discussions. This work
is supported in part by funds provided by the U.S. Department of
Energy (DOE) under the cooperative research agreement
DEFG02-05ER41360. DG has been supported by the Marie Curie
Fellowship MC-OIF 021421 of the European Commission under the
Sixth EU Framework Programme for Research and Technological
Development (FP6). DG would like to thank the Erwin-Schr{\"o}dinger International Institute for Mathematical Physics (ESI) for the hospitality during the final preparations of this manuscript.


\section*{References}

\providecommand{\href}[2]{#2}\begingroup\raggedright\endgroup

\end{document}